\documentclass[twocolumn,showpacs,aps,prl,superscriptaddress]{revtex4}

\usepackage{graphicx}
\usepackage{amsmath}
\usepackage{latexsym}

\begin{document}

\title{Tilt Induced Localization and Delocalization in the Second Landau Level}

\author{G.A. Cs\'athy}
\affiliation{Department of
Electrical Engineering, Princeton University, Princeton, NJ 08544}
%\email{gcsathy@princeton.edu}

\author{J.S. Xia}
\affiliation{University of Florida, Gainesville, FL 32611}
\affiliation{National High Magnetic Field Laboratory, Tallahassee,
FL 32310}

\author{C.L. Vicente}
\affiliation{University of Florida, Gainesville, FL 32611}
\affiliation{National High Magnetic Field Laboratory, Tallahassee,
FL 32310}

\author{E.D. Adams}
\affiliation{University of Florida, Gainesville, FL 32611}
\affiliation{National High Magnetic Field Laboratory, Tallahassee,
FL 32310}

\author{N.S. Sullivan}
\affiliation{University of Florida, Gainesville, FL 32611}
\affiliation{National High Magnetic Field Laboratory, Tallahassee,
FL 32310}

\author{H.L. Stormer}
\affiliation{Columbia University, New York, NY 10027}
\affiliation{Bell Labs, Lucent Technologies, Murray Hill, NJ 07974}

\author{D.C. Tsui}
\affiliation{Department of
Electrical Engineering, Princeton University, Princeton, NJ 08544}

\author{L.N. Pfeiffer}
\affiliation{Bell Labs, Lucent Technologies, Murray Hill, NJ 07974}

\author{K.W. West}
\affiliation{Bell Labs, Lucent Technologies, Murray Hill, NJ 07974}

\date{\today}

\begin{abstract}

We have investigated the behavior of electronic phases 
of the second Landau level under tilted magnetic fields. 
The fractional quantum Hall liquids 
at $\nu=2+1/5$ and $2+4/5$ and the solid phases
at $\nu=2.30$, 2.44, 2.57, and 2.70
are quickly destroyed with tilt. 
This behavior can be interpreted as a tilt driven localization
of the 2+1/5 and 2+4/5 fractional quantum Hall liquids and a 
delocalization through melting of solid phases in the top Landau level,
respectively. The evolution towards the classical Hall gas of the
solid phases is suggestive of antiferromagnetic ordering.

\end{abstract}
\maketitle

Two dimensional electron systems (2DES) 
subjected to perpendicular magnetic fields $B$ exhibit a myriad 
of ground states. Perhaps the most well known of these are
the incompressible states called the integer (IQHL) and
fractional quantum Hall liquids (FQHL) \cite{fraq}.
The low frequency transport signature of these IQHLs and FQHLs
is the quantized Hall resistance
R$_{xy}$ accompanied by a vanishing diagonal resistance R$_{xx}$.
While the IQHLs are the consequence
of purely single particle physics, the
FQHLs forming at certain fractional values of the Landau
level filling factor $\nu$ can only be explained by considering
interparticle interactions \cite{laughlin,jain}. 
The various series of FQHLs are successfully accounted by
the composite fermion theory \cite{jain}.

Strong interparticle interactions give rise to a second
class of many particle ground states: that of compressible solids.
With the availability of samples with continuously improving quality
a number of solid phases have been found \cite{WS,WSrec,WSint,bubble}. 
First examples are the high field insulating and reentrant insulating
phases of the lowest LL at the highest $B$-fields \cite{WS},
phases that have been associated with the Wigner solid (WS) \cite{Wigner}. 
A recent work in the highest quality samples available 
found that there are two types of WS phases in this regime
\cite{WSrec}. Microwave resonances close to $\nu=1$, 2, 3, and 4
have also been interpreted as being due to the WS \cite{WSint}.
Other examples of solid phases are the recently discovered
electronic stripe and bubble phases in high Landau levels
also referred to as charge density waves (CDW) \cite{bubble}.
While the transport signature of stripes is anisotropic,
that of the bubble phases is isotropic and it is described by the
reentrant integer quantum Hall (RIQH) effect \cite{bubble}.
The RIQH effect is manifest in a R$_{xy}$ quantized to an integer
multiple of the quantum Hall resistance combined with a vanishing R$_{xx}$ 
but which, unlike the IQHE, is centered at a filling factor that is 
different from the integer value to which the R$_{xy}$ plateau is quantized.
This behavior is a consequence of a disorder pinned
solid phase forming in the top LL when multiple LLs are occupied. 
While substantial progress has been made, the nature of
the CDW phases has not yet been fully understood.

The 2nd Landau level is very special
being at the borderline of the two very different regimes \cite{2nd,xia}.
On one hand the lowest LL
is dominated by FQHLs, a series 
of phases that is terminated on the low filling side by
the high field WS. On the other hand beyond the 2nd LL
stripe and bubble phases dominate and, with the exception of a narrow
temperature range $80 \lesssim T \lesssim 120$~mK \cite{gervais},
there are no FQHLs present.
Thus in the 2nd LL the FQHL phases of the lowest LL and the
CDW phases of high LLs are expected
to compete leading to an intricate behavior. 
Indeed, the $\nu=3+1/5$, 3+4/5, 2+1/5, 2+1/3, 2+2/3, 2+4/5, 2+2/5, and
possibly 2+2/7 FQHLs as well as eight RIQH states
have been reported in the 2nd LL \cite{2nd,xia}. 
Besides the alternating FQHL and RIQH states,
there are special fingerprint FQHLs at even denominator 
filling that are present in the 2nd LL only.
These states at $\nu=5/2$, 7/2 \cite{2nd,xia,5/2} and 
possibly at 2+3/8 \cite{xia} are believed to arise
from a BCS-like pairing of composite fermions \cite{read}.
The evolution with tilted magnetic fields
of the $\nu=5/2$ and 7/2 states toward anisotropic
states that are very similar to stripes of half-filled higher LLs 
\cite{5/2tilt} and the presence of the RIQH states between the FQHLs
are regarded as evidence of the delicate balance between the phases
of the 2nd LL. While it has been suggested \cite{2nd,xia} 
and theoretically independently obtained \cite{Btheory}
that the RIQH states of the 2nd LL are isotropic collective
insulators also called bubble phases, an independent experimental 
verification is still lacking.

In this Letter we  have investigated the influence of a magnetic 
field parallel to the confinement plane of the 2DES
on the various electronic phases in the 2nd LL
in the $2<\nu<3$ range using the tilted
field technique. We find that  the recently discovered 
RIQH states are rapidly destroyed with tilt. 
Such a behavior is not consistent
with single particle localization in the top LL, therefore it
constitutes experimental evidence
that the RIQH states in the 2nd LL are collectively pinned insulators. 
The rapid evolution with tilt of
R$_{xy}$ of the RIQH states from the values
of the nearby integer plateaus towards the classical Hall 
value can be interpreted as melting of this collective
phase into a classical Hall gas. 
Furthermore, since tilting changes the ratio of the Zeeman
and cyclotron energies, the data
suggests that spin interaction plays an important role in the
formation of these collective phases. We surmise
that the RIQH phases are not fully spin-polarized but
have substantial antiferromagnetic order. 
These phases could be the
first examples of antiferromagnetically ordered
solids in a single layer 2DES in the quantum Hall regime.
In addition, the well
developed $\nu=2+1/5$ and 2+4/5 FQHLs are found to be
driven insulating while the 2+1/3
and 2+2/3 states survive to the largest tilt angles we can reach.

The low frequency magnetoresistance measurements were performed
on a $\delta$-doped $30$nm wide GaAs/AlGaAs quantum well
at an excitation current of 1~nA.
The 2DES has been prepared with a brief illumination with
a red LED at low temperatures and has
a density of $3.0\times 10^{11}$cm$^{-2}$ and an
exceptionally high mobility of $2.7\times 10^7$cm$^2$/Vs.
The challenging task of cooling to mK temperatures and in-situ tilting
in this low temperature environment is achieved in a special
hydraulically driven rotator \cite{xiaRot} equipped with
sintered Silver heat exchangers immersed in a $^3$He bath \cite{5/2}. 

The diagonal and off-diagonal resistances as function of the
magnetic field component perpendicular to the 2DES $B_{\bot}$ at a set of 
representative tilt angles $\theta$ are shown in Fig.1.
The bath temperature is 9~mK.
The traces in purely perpendicular field or at
$\theta=0^{\circ}$ are located in the middle of Fig.1. These traces 
are similar to those of samples of comparable parameters \cite{2nd,xia}. 
At $\theta=0^{\circ}$ we observe the
FQHLs at $\nu=2+1/5$, 2+1/3, $2+2/3$, and 2+4/5 as well as
the well-developed $\nu=5/2$ state. 
In the vicinity of $\nu \simeq 2.30$, 2.44,
2.57, and 2.70 the RIQH states develop since
the vanishing R$_{xx}$ is accompanied
by an R$_{xy}$ that jumps to the nearest integer value.
In between the  FQHL and RIQH states,
R$_{xy}$ follows the classical Hall line. 
We note that in a recent experiment the $\nu=2+1/5$ FQHL
is well developed at 40~mK but it
started to evolve toward a localized phase at 16~mK 
and the RIQH phase around $\nu=2.44$ is interrupted
by the developing $2+2/7$ FQHL \cite{xia}.

\begin{figure}[t]
\begin{center}
\includegraphics[width=3.8in]{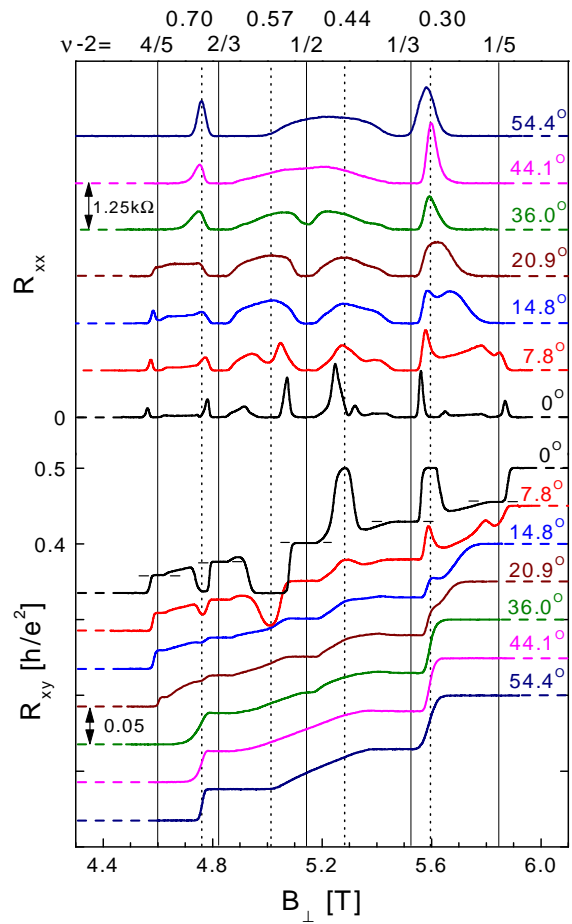}
\end{center}
\caption{\label{f1}
Dependence of R$_{xx}$ and R$_{xy}$ on $B_{\bot}$ at various tilt
angles $\theta$ measured at 9~mK. Filling factors are shown on the top
scale. Horizontal lines highlight the
plateaus of R$_{xy}$ of the FQHLs at $\theta=0^{\circ}$.
}
\end{figure}

As shown in Fig.1, with increasing tilt
the $2+1/3$ FQHL stays robust while
the $2+1/5$ FQHL quickly vanishes. Once destroyed,
the 2+1/5 FQHL does not reemerge with further tilting.
This behavior is summarized in Fig.2, where we have plotted the 
tilt dependence of R$_{xy}$ at filling factors at which FQHLs develop.
None of these FQHLs display the cusp in R$_{xx}$ that is
associated with spin transitions \cite{spinpol,spin2/3}. 
Our data therefore suggests that the 2+1/3 and 2+1/5 FQHLs are
spin polarized, just as their $\nu=1/3$ and 1/5 counterparts are 
in the lowest LL \cite{laughlin}. 
It is interesting to note that similar tilts could lead
to a spin-unpolarized state for the $\nu=2+2/3$ FQHL
since the $\nu=2/3$ FQHL in the lowest LL
has convincingly been shown to be spin-unpolarized \cite{spin2/3,freytag}.
However, up to the largest tilt angle $\theta =54.4^{\circ}$ of
our experiment we do not observe any sign of a spin transition. We explain this
behavior by either a $B$-field that is
too small to polarize this state or,
more likely, a fully spin-polarized $\nu=2+2/3$ FQHL. 
At negligible spin mixing
a fully spin-polarized 2+2/3 state can be derived 
from the $\nu=2+1/3$ spin-polarized FQHL
by particle-hole symmetry within the top LL.

\begin{figure}[t]
\begin{center}
\includegraphics[width=3.6in]{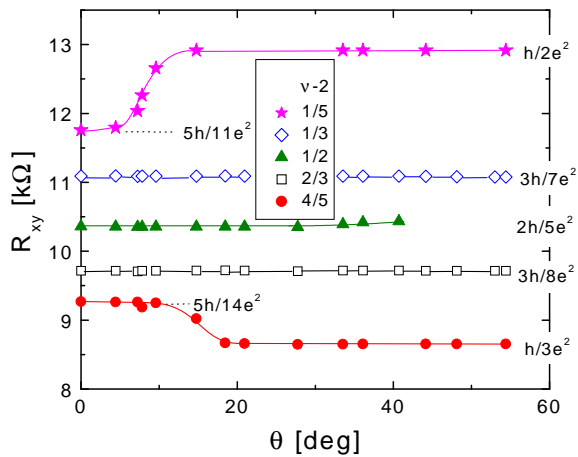}
\end{center}
\caption{\label{f2}
Tilt dependence of R$_{xy}$ at $\nu=2+1/5$, 2+1/3, 2+1/2, 2+2/3, and 2+4/5
measured at 9~mK. While R$_{xy}$ for the
2+1/3 and 2+2/3 FQHLs is unchanged, for
the 2+1/5 and 2+4/5 FQHLs it evolves from $h/\nu e^2$
toward the nearest integer quantum Hall value.
}
\end{figure}

We have seen that the $\nu=2+1/5$ and 2+4/5 FQHLs turn insulating with tilt.
A similar localization transition has been recently observed in
the lowest LL in a low density 2D hole sample in which tilt localizes
the terminal FQHL with $\nu=1/3$ \cite{panTilt}. One route to
localization in the top LL with effective filling 1/5 is the
enhanced surface roughness scattering.
This scattering mechanism is due to the
single particle wavefunction being squeezed in tilted field
against the interface of the confining potential.
We think, however, that such a scenario is unlikely. At the
approximately
$15^{\circ}$ tilt at which these FQHLs are destroyed the center of the
single particle wave function in a 30~nm quantum well 
does not shift substantially. 
A second route that renders
the FQHL localized and that most likely explains our data
is given by the evolution with tilt
of the ground state from the FQHL toward a pinned collective insulator. 
Such a transition occurs if an electronic solid becomes energetically
favored at large tilt angles. 

We focus next on the evolution of the RIQH states
with tilt. At small tilt angles the RIQH phase 
at $\nu=2.70$ and at constant $T$ gradually weakens with tilt, 
a response briefly mentioned in Ref.\cite{2nd}.
As the tilt increases we find that R$_{xy}$ evolves from $h/3e^2$, the
value of the nearby $\nu=3$ IQH plateau, and
reaches a value very close to the classical Hall
value beyond $\theta \simeq 10^{\circ}$ tilt. This behavior, together with the
results for the other RIQH states, is summarized in Fig.3a.
The observed behavior is interpreted as a 
delocalization transition under tilt.
Since for localized single particles there is no known mechanism for
delocalization driven by the application of a magnetic field parallel
to the 2DES, the reentrant behavior cannot be due to single particle
localization. The tilt induced destruction of the
RIQH behavior therefore demonstrates a collectively pinned
solid at $\theta=0^{\circ}$. 

\begin{figure}[b!]
\begin{center}
\includegraphics[width=3.6in]{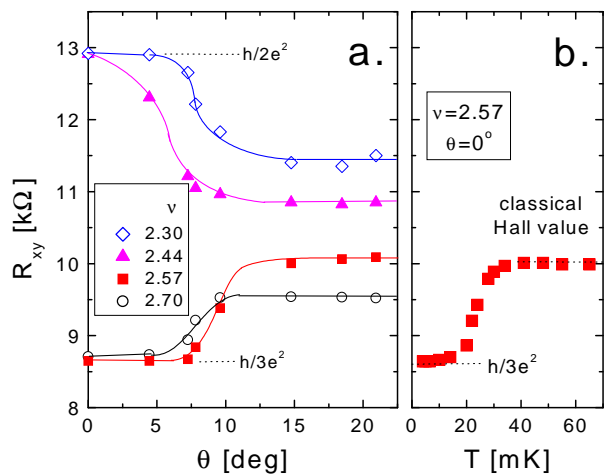}
\end{center}
\caption{\label{f3}
Angle dependence of R$_{xy}$ at $\nu=2.30$, 2.44, 2.57, and 2.70
at 9~mK (panel a.) and
$T$ dependence at $\nu=2.57$ and at zero tilt (panel b.).
Lines are guides to the eye.
}
\end{figure}

With increasing tilt, the $\nu=5/2$ state weakens and
beyond $\theta \simeq 44^{\circ}$ cannot be discerned any more.
It has been proposed that
this is due to a symmetry breaking mechanism induced
by the parallel component of the $B$-field \cite{5/2tilt}. 
As a result, the isotropic $\nu=5/2$ state at zero tilt becomes an
anisotropic phase similar to the stripe phases in higher
LL. We have investigated
if a similar symmetry breaking occurs for the RIQH states
by running the current both parallel and perpendicular to the
direction of the parallel component of the $B$-field.
In doing so
there is little change in the data (not shown).
We conclude that, similarly to the $\theta=0$ case, in tilted fields
there is no anisotropic behavior
except in the vicinity of the $\nu=5/2$.

The rapid transformation with tilt of the RIQH states into 
the classical Hall gas can be interpreted as a tilt-induced
melting of the bubble phase. The evolution at $\nu=2.30$, 2.44,
2.57, and 2.70 of R$_{xy}$
from the nearby integer plateau toward the classical Hall value
cannot be explained by
a transition from the bubble phase to a phase of singly localized
particles. This is consistent with the earlier observation
that single particle localization due the enhanced
interface roughness scattering is most likely not 
substantial for the FQHLs.
Thus, the destruction of the RIQH states with tilt
is most likely not disorder-driven. 
A second, intriguing possibility 
that can explain the behavior with tilt
is that the bubble phase is only partially spin polarized.
We consider such a scenario because
a fully spin polarized solid is not expected to be affected by tilting.
Melting of the partially polarized bubble phase can occur
if its energy becomes higher than that of
the classical Hall gas as the tilt angle increases.
The abrupt $T$-dependence of R$_{xy}$, shown in Fig.3b for $\nu=2.57$
is not inconsistent with such a scenario.   
A partially spin-polarized electronic solid
most likely has antiferromagnetic ordering.

The exchange interaction modeled by the Heisenberg Hamiltonian
for the spin 1/2
is an essential ingredient in understanding magnetic properties
of quantum solids. 
Magnetism in the $B=0$ WS was found to be
determined by the competition of different ring
exchange processes \cite{AFWC}.
Even and odd circular permutations lead to 
ferromagnetic and antiferromagnetic coupling, respectively 
and the type of lattice can influence the dominant term \cite{AFWC}. 
A known experimental
realization of a 2D quantum solid of spin 1/2 particles
is the solidified 2nd layer $^3$He prepared on a 
graphite substrate \cite{3He}.
Since the single particle wavefunctions of electrons in the top
LL have significant overlap, in our system there is a considerable
exchange. We speculate that the 
exchange has a dominant antiferromagnetic term
in the bubble phases of the 2nd LL.

Finally we note, that at tilt angles beyond $30^{\circ}$, 
as the IQH plateaus extend over the $2+1/5$ and $2+4/5$ fillings,
any sign of reentrance has disappeared and $\nu=2.30$ and $2.70$
become the demarcation line between the 2 and 2+1/3
and the 3 and 2+2/3 plateaus, respectively.
This results, as shown in Fig.1, in pronounced peaks in R$_{xx}$ and
a steep R$_{xy}$ as function of $B_{\bot}$
close to $\nu=2.70$ and 2.30. We note that 
due to this steep R$_{xy}$, its value
at $\nu=2.30$ and 2.70 has a substantial error 
propagating from small errors of the tilt angle.
At the highest tilt angle $\theta > 48^{\circ}$ we observe
an asymmetry between the plateaus of the 2+1/3 and 2+2/3 FQHLs
that we do not understand. Using $B_{\bot}$ as abscissa,
the plateau length of the 2+1/3 FQHL shrinks and
that of the 2+2/3 FQHL grows.
In fact at $\theta=54.4^{\circ}$, as shown in Fig.1, 
the plateau of the 2+2/3 FQHL
will extend beyond $B_{\bot}$ corresponding to $\nu=2.57$.

To summarize, we found that an interesting tilt and filling factor dependent
interplay of localization and delocalization
shapes the dc transport in the 2nd LL. These transitions 
are a result of the delicate balance of various phases of the 2nd LL.
We think that an important 
consequence is that the 2+1/5 FQHL cedes its place
as a ground state to a collective solid, while
the collectively pinned solids associated with reentrance 
of the integer quantum Hall plateaus
melt into a classical Hall gas.
The RIQH states at zero tilt are electronic solid phases 
with possible antiferromagnetic order.

We acknowledge F.D.M. Haldane for insightful discussions.
The work at Princeton was supported by DOE and NSF.
The work at Columbia was supported by DOE, NSF and by the
W.M. Keck Foundation. The experiment was
carried out in the high $B/T$ facility of the NHMFL
in Gainesville, which
is supported by NSF Cooperative Agreement DMR-0084173 and 
by the State of Florida.

\end{document}